\documentclass[twocolumn,showpacs,amsmath,amssymb,floatfix]{revtex4}

\usepackage{graphicx}
\usepackage{dcolumn}
\usepackage{bm}

\begin{document}

\title{Bell correlations and equal time measurements}

\author{Frederick H.~Willeboordse}
\homepage{http://chaos.nus.edu.sg/}
\email{phyfhw@nus.edu.sg, willeboordse@yahoo.com}
\affiliation{Department of Physics, The National University of Singapore, Singapore 117542}

\date{\today}

\begin{abstract}
According to the Bell theorem, local hidden variable theories cannot reproduce all the predictions of quantum mechanics. An important consequence is that under physically reasonable assumptions quantum mechanics predicts correlations that seem impossible to obtain from a realistic system. In this paper, two simple binary apparatuses are discussed that can accurately mimic correlations predicted by quantum mechanics \textit{if} the correlations are determined by a coincidence measurement as is commonly done. It is argued that in order to exclude local hidden variables, coincidence monitors should be avoided.
\end{abstract}


\maketitle



\section{Introduction}

A key question regarding the foundation of quantum mechanics is whether or not it can be considered a statistical representation of an underlying realistic theory. In this context, realism refers to the existence, before measurement, of physical states described by so-called 'hidden variables'. In 1935, Einstein, Podolsky and Rosen (EPR)\cite{EPR} argued that correlations between entangled particles imply the necessity of such hidden variables. In 1964, John S.~Bell then described a modified EPR setup which would make it possible to experimentally determine whether or not hidden variables could reproduce the predictions of quantum mechanics\cite{Bell-1964}. A large number of experiments has since been carried out and all these experiments confirm the quantum mechanical predictions to a very high degree thus apparently ruling out hidden variable theories\cite{Shimony_Stanford}. Although loopholes cannot entirely be discounted, considering the variety of the experiments, it seems extremely unlikely that experimental flaws can account for the results obtained. The Bell theorem stating that ``no physical theory of local hidden variables can ever reproduce all of the predictions of quantum mechanics'' therefore appears to stand on firm ground and a realistic theory impossible.

 
 

In this paper, the following question is addressed: Is it possible to reproduce Bell correlations from a locally realistic perfectly efficient classical apparatus \textit{if} the correlations are calculated in the same fashion as in the experiments that violate the Bell inequalities? That is to say when coincidence monitors are used or equal time measurements taken.

In the literature, Bell inequality generally stands for a variety of inequalities that can be used to discriminate between the predictions of quantum physics and possible realistic models. A violation of a Bell inequality is thought to only be possible in a quantum mechanical description. Closely related to the Bell inequalities are the Bell correlations which are such that they violate the Bell inequalities and hence are a tell-tale sign of a non-realistic system. Lastly, Bell experiments are experiments that test the Bell inequalities.

\section{Experimental setup}
\begin{figure}[htb]
   \includegraphics*[width=6cm]{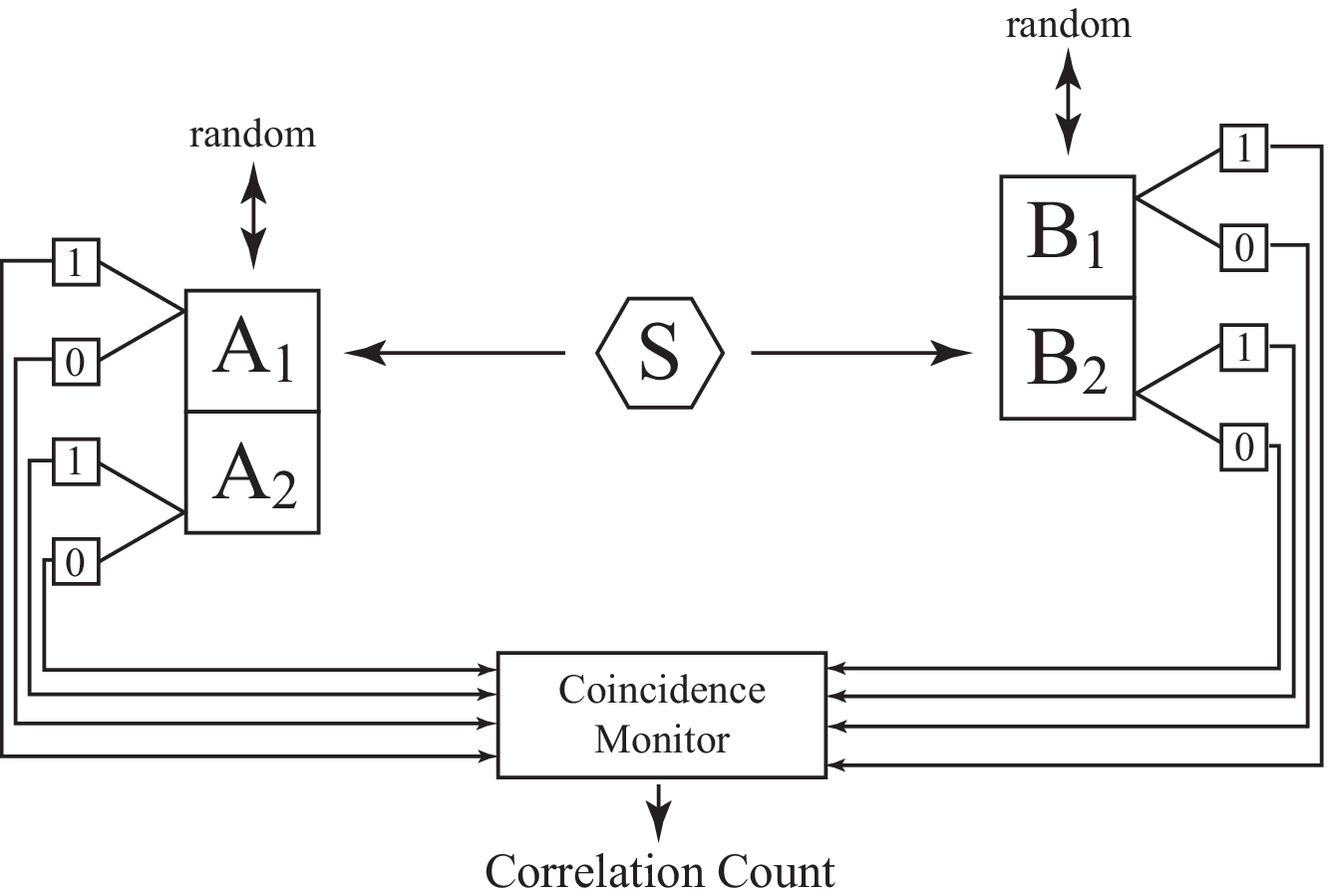}
   \caption
   {
     Typical setup in a CHSH style Bell correlation experiment. The source S sends out a pair of entangled particles toward detectors A and B that can randomly be in 
     position 1 or 2. The binary output of the detectors is collected by a coincidence monitor. 
   }
   \label{figure1}
\end{figure}

Let us consider the Bell inequalities due to Clauser, Horne, Shimony and Holt (CHSH)\cite{CHSH}. A typical experimental setup is depicted in Figure \ref{figure1}. In this setup, a source creates an entangled pair of particles (usually photons) that interact with a detector that can be in one of two possible settings (e.g.~the direction with respect to which polarization is measured). The detector setting must be completely random, unknown to the source at the time a pair of particles is created, and independent of the other detector's setting. Consequently, communication between the detectors, the source and the detectors etc.~is not allowed thus ensuring that the physical processes occurring when a particle hits a detector are local. When a particle interacts with a detector, a binary outcome is generated which together with the detector setting is either forwarded to a coincidence monitor or recorded with an accurate clock for later comparison (the latter is conceptually equivalent and not further discussed here).

One can then determine the various (anti-)correlations between the right and left hand detector settings and compare the experimental outcomes with quantum mechanical and classical calculations. Denoting the expectation values for the (anti-)correlations as $E_{\phi}(A_i,B_j)$ where $\phi$ is the angle between the detectors $A$ and $B$ that have settings $i,j = 1,2$ respectively, CHSH consider the following sum of (anti-)correlations \cite{CHSH}
\begin{equation}
  S_{\phi} = E_{\phi}(A_1,B_1) + E_{\phi}(A_1,B_2) + E_{\phi}(A_2,B_1) - E_{\phi}(A_2,B_2).
\end{equation}
Quantum mechanically, the maximum is $S_{\phi} = 2\sqrt{2}$ while in a realistic model, the maximum is $S_{\phi} = 2$ \cite{Shimony_Stanford}. The realistic result is easily illustrated: assume that the photon carries the binary information for each possible detector setting. For example, if we take $A_1=1$, $A_2=0$, $B_1=0$ and $B_1=0$ (in total there are 16 possibilities), then $S = 2$; changing a 0 to a 1 may at best improve the anti-correlation with one detector but this will be at the expense of the anti-correlation with another detector. Therefore, if an experiment produces an $S_{\phi}$ clearly larger than 2, a realistic explanation appears impossible. And this is exactly what was demonstrated in sophisticated experiments by e.g.~Aspect\cite{Aspect-1982}, Freedman\cite{PhysRevLett.28.938} and many others where $\mathrm{Max}( S_{\phi}) > 2$ was found.

Although it seems that conceptually, the existence of a coincidence monitor does not change the logic of the argument, this is only so if from the outset the current quantum mechanical interpretation is taken to be correct. But if the the purpose is to verify this interpretation, then it cannot be taken as a matter of course that a coincidence monitor is included in the setup or that equal time measurements are carried out. It is therefore investigated here what the possible impact of a coincidence monitor is by considering two classical apparatuses that include this experimental component.

\section{Apparatus 1}

Here, instead of viewing it as an experimental setup with photons, we take Figure \ref{figure1} to depict a classical apparatus that functions as follows:
\begin{itemize}
  \item The source sends out two bullets in opposite directions. Randomly, one of the bullets carries a 1 and the other a 0.
  \item During the flight, randomly and independently from each other, the detectors are set into position 1 or 2.
  \item If the left detector is in position $A_1$, randomly, it either outputs the bit carried by the bullet immediately when the bullet arrives, or a NOT of the bit           with a given delay $\Delta t$ after the bullet arrives.
  \item If the left detector is in position $A_2$, randomly, it either outputs the bit carried by the bullet immediately when the bullet arrives, or with a given 
        delay $\Delta t$ after the bullet arrives.
  \item If the right detector is in position $B_1$, it outputs the bit carried by the bullet immediately when the bullet arrives.
  \item If the right detector is in position $B_2$, it outputs the NOT of the bit with a given delay $\Delta t$ after the bullet arrives.
  \item All actions are local. There is no communication between the detectors.
\end{itemize}

\begin{figure}[htb]
   \includegraphics*[width=6cm]{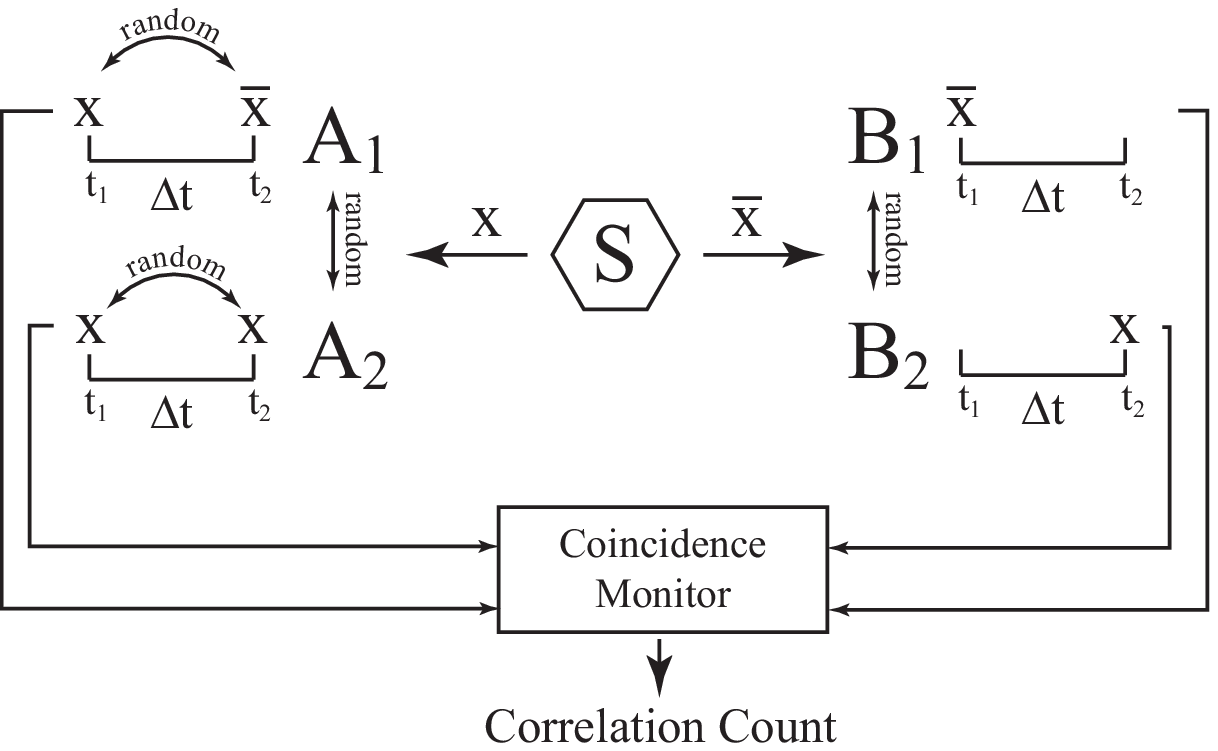}
   \caption
   {
     Schematic overview of the functioning of apparatus 1. The binary variable x randomly takes on the value 0 or 1.
   }
   \label{figure2}
\end{figure}

These instructions are summarized in Figure \ref{figure2}.

When calculating the correlations on an event basis, as expected, this apparatus will not give Bell correlations. However, if the correlations are calculated, just like in actual experiments, with the help of coincidence monitoring, Bell correlations are obtained. Indeed, the apparatus will yield the maximum possible $S=3$. 

\begin{figure}[htb]
   \includegraphics*[width=4cm]{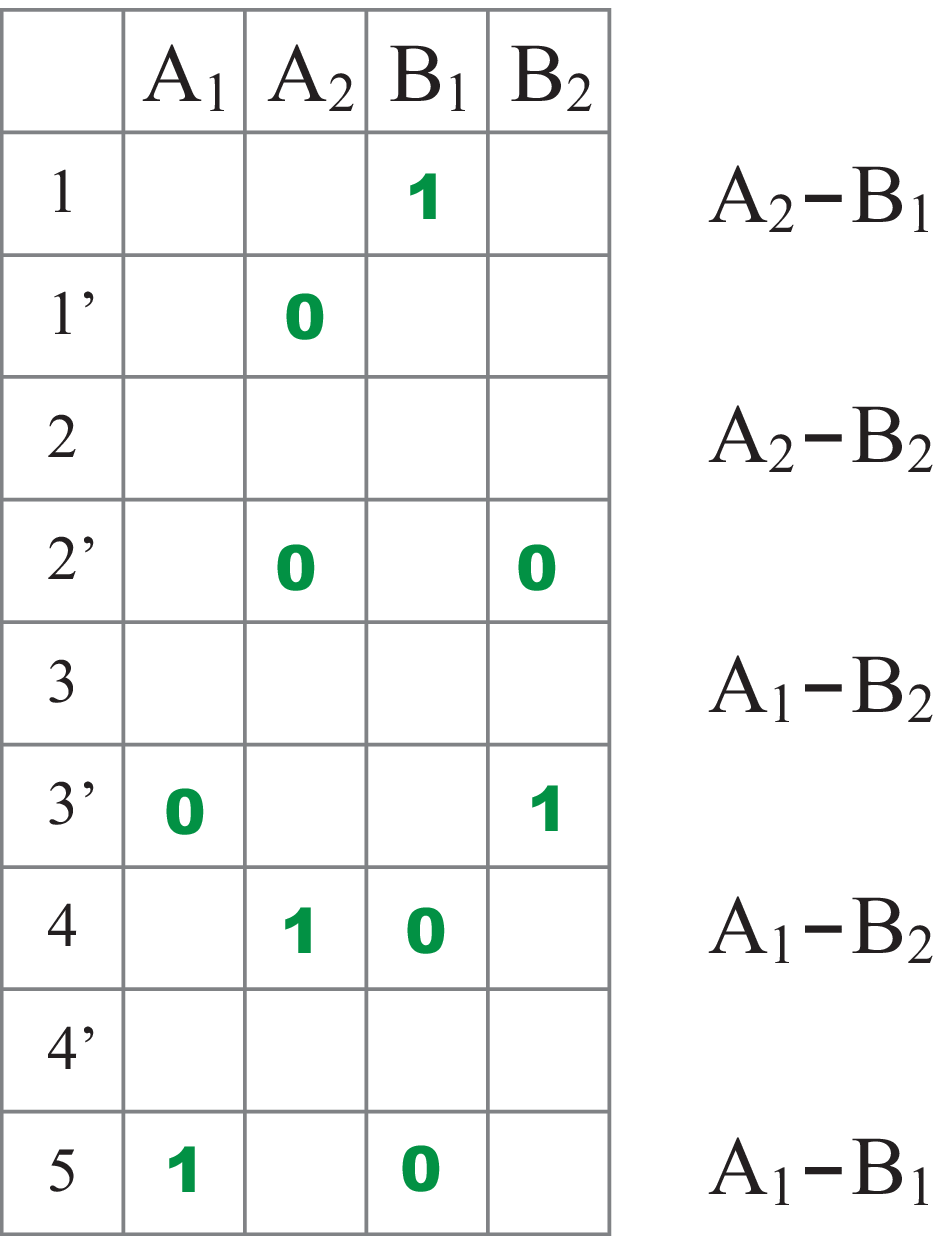}
   \caption
   {
     Possible output of apparatus 1. The numbers on the left indicate which bullet the output stems from, and the prime indicates an output that occurs after a delay
     of $\Delta t$.
   }
   \label{figure3}
\end{figure}

Figure \ref{figure3} shows some possible data of a simulation as they arrive at the coincidence monitor. There is a 50\% chance that an output of one detector is at a different time of the other detector, giving single counts 66\% of the time that are ignored by the coincidence monitor. Experimentally, such single counts would be very hard to distinguish from dark counts but if the Bell theorem holds, all outputs should be coincidence counts. Hence in principle, an apparatus like this can be excluded with a perfect experimental setup having extremely high efficency and, in the case of photon experiments, very few dark counts (this seems to exclude all photon experiments to date). While the efficency in an ion experiment such that by Rowe \cite{Rowe-2001} is extremely high, measurement is carried out by synchronized detection lasers hence builiding in coincidence at even earlier stage. Such experiments are therefore not suitable to exclude a mechanism as in the described apparatus.

Statistically, the apparatus reproduces experimental outcomes well. The probability of obtaining a 1 or 0 from all detector settings is 50\%, and the probabilities of each of the detector combinations is 25\%.

It may seem that what is exploited here is either the detection or the fair sampling loophole. While one can of course take that viewpoint, I would argue that this is not the case.

\begin{itemize}
  \item \textbf{Detection loophole}. All events are recorded and nothing is dropped at the detection stage. For every bullet there is an output at each side. There is therefore
       no detection loophole in the apparatus described.
  \item \textbf{Fair sampling loophole} This loophole seems to have slightly differing definitions in the literature. Here the position is taken that fair 
       sampling means that 
       the experimental data are a fair reflection of what would be obtained if all the experiments components were perfect. Therefore, given that the coincidence
       monitor is part of the experiment, the experimentally obtained data appear to be fair samples. By the same token, however, the apparatus delivers fair data
       as well.
       
       Now one could argue that if the experimental setup was perfect then there would not be any single counts. However, this is something inferred from quantum 
       mechanics, and thus an assumption to be determined experimentally, and not something that can serve an input to fair sampling. Consequently, in my view, the
       apparatus does not violate the fair sampling assumption.
       
       In other words, whether or not to include a coincidence monitor or to carry out equal time measurements is an experimental choice. Given the choice, 
       the data are fair but whether the choice is fair needs to be justifiable without resorting to the outcome expected from quantum mechanics.
\end{itemize}


Although the above shows that the Bell inequalities can easily be violated with the help of a coincidence monitor, it is nevertheless illustrative to show that the apparatus can mimic the quantum mechanical prediction for S. If we take the detectors to be polarizers that can be rotated with respect to each other and for simplicity fix the angles between $A_1$ and $A_2$, and $B_1$ and $B_2$ respectively to 90 degrees, then the quantum mechanical $S$ is given by \cite{Shimony_Stanford}:
\begin{equation}
  S_{\phi} = 2 \cos(2\phi) + 2 \sin(2\phi).
  \label{eq:sqm}
\end{equation} 

\begin{figure}[htb]
   \includegraphics*[width=6cm]{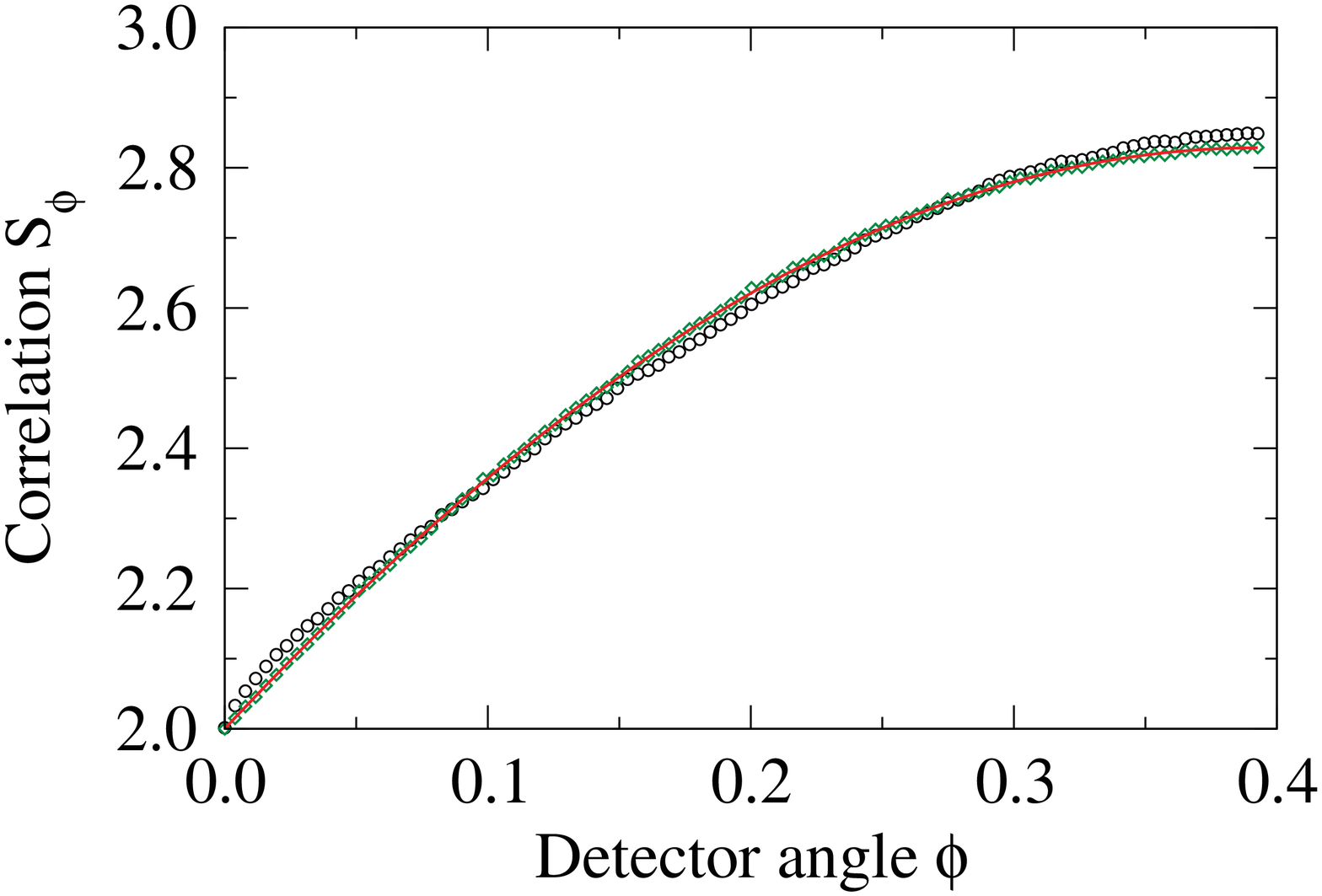}
   \caption
   {
     Comparison of the quantum mechanical prediction (red solid line) and the output from the apparatuses. The diamonds (green) show the out put from apparatus 1 
     and are basically right on top of the line. The circles (black) show the output of apparatus 2. 
   }
   \label{figure4}
\end{figure}

Figure \ref{figure4} shows a comparison between the dependence on $\phi$ in Eq.~\ref{eq:sqm} and the apparatus where the choice of whether or not to apply a time delay to the output of $B_2$ was subjected to the following  $\phi$-dependent probability
\begin{equation}
  p_{\phi} = 2 \sqrt{2} \sin(2 \phi + \frac{\pi}{4}) - 2.
\end{equation}

As can clearly be seen, the data match extremely well. It should be stressed though that it is unlikely that there is a fundamental meaning to this. The aim here is to simply show that it can be done and not to develop a possibly physically meaningful model.

Next a second similar apparatus is presented. The main difference being that information from successive bullets is temporarily stored in the detectors. This kind of procedure reduces the number of single counts. Furthermore, it may make long time delays a bit more palatable. As such, apparatus 2 does not add much but it does illustrate that all kinds of apparatuses of the kind discussed can be constructed.


\section{Apparatus 2}
The design of apparatus 2 is as follows:
\begin{itemize}
	\item The source sends out two bullets in opposite directions. The bullets carry with them the outputs for the possible detector settings.
	\item The bullets alternately carry slightly different instructions (instruction sets i and ii).
	\item During the flight of the bullets, the detectors are randomly and independently set to position 1 or 2. There is no communication between the detectors or the detectors and
	      the source at any time.
	\item The detectors obtain the instructions from the bullets and output a 0 or 1
  \begin{itemize}
	  \item Instruction set i: $A_1 = 0, A_2 = 1, B_1 = 0, B_2 = 1$. Detector settings $A_1$ and $B_2$ generate the output immediately. Detector settings 
	        $A_2$ and $B_1$ store the instruction and output it the next time a bullet arrives.
	  \item Instruction set ii: $A_1 = 1, A_2 = 0, B_1 = 1, B_2 = 0$. Detector setting $A_1$ generates the output immediately. Detector settings $A_2$ and $B_2$ store
	        the instruction and output it with a slight delay. Detector setting $B_1$ stores the instruction and outputs it when the next bullet arrives.
  \end{itemize}
  \item The output is forwarded to the coincidence meter, or alternatively recorded with an accurate time stamp.
\end{itemize}

\begin{figure}[htb]
   \includegraphics*[width=7cm]{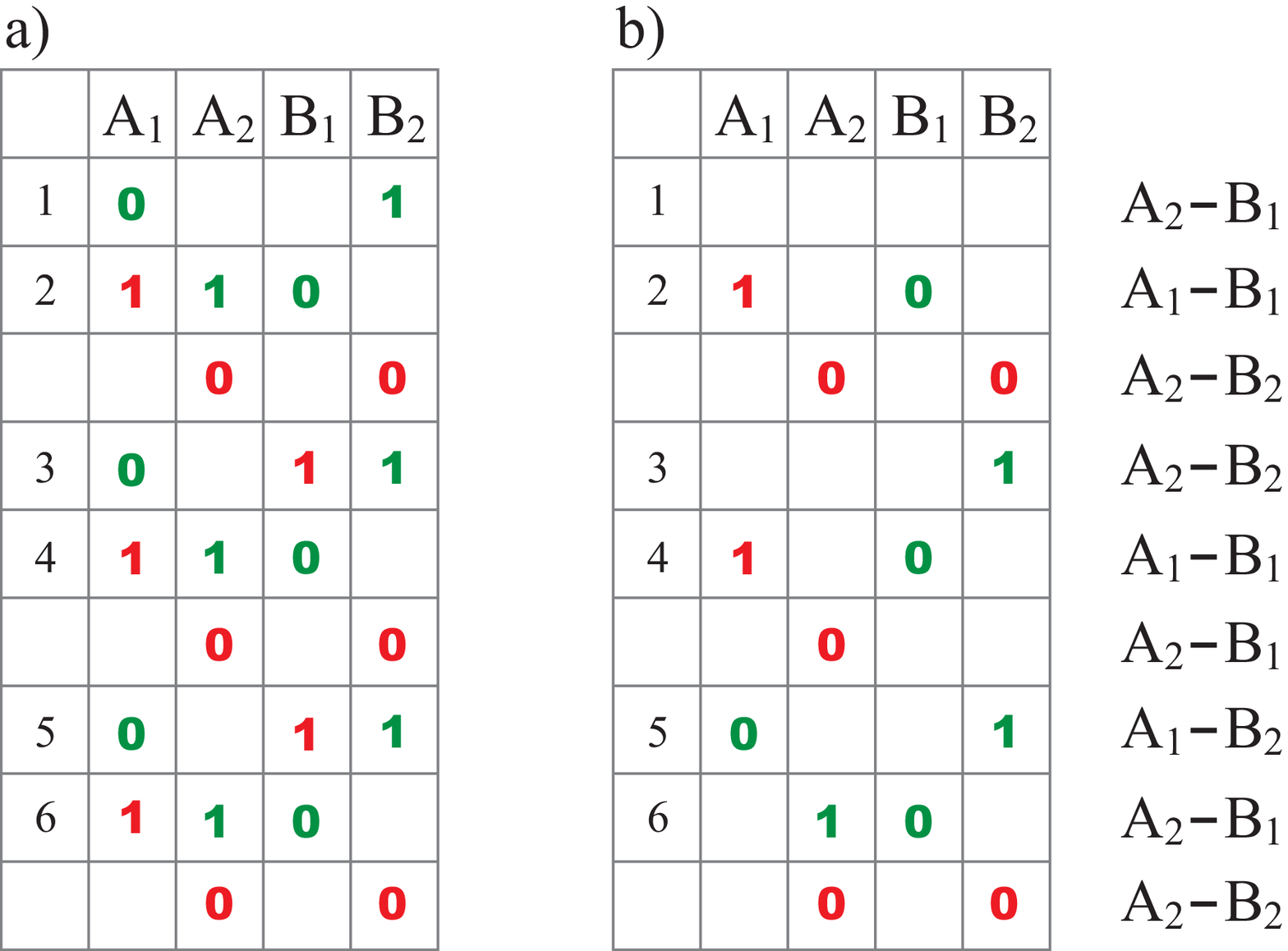}
   \caption
   {
     a) Logically attainable instructions for six consecutive bullets. b) A possible set of data as it is received by the coincidence meter. Instruction set i) is
     colored green and instruction set ii) colored red. The numbers on the left in a) indicate which bullet arrives at the detector.
   }
   \label{figure5}
\end{figure}

Figure \ref{figure5} shows the instructions for the four possible detector settings containing all the logically attainable combinations for six consecutive bullets in a), while a possible sequence of outcomes as forwarded to the coincidence meter due to the randomly switching detectors is shown in b). It should be noted again, that the setup exactly conforms to requirements for Bell experiments. There is no communication between the detectors and no events are dropped. Not surprisingly then, instructions i and ii do by themselves not violate the Bell inequalities and S = 2.


If we now take the data from Figure \ref{figure5} and carry out the coincidence measurement exactly as it is done in experiments, $S = 3$ is obtained. The reason is easy to understand, due to the unequal delays in the output, zeros or ones belonging to different bullets are compared. By subjecting the output of $B_2$ to the ad-hoc $\phi$-dependent probability
\begin{equation}
  p_{\phi} = 2.08 \left( \sin(2 \phi + \frac{\pi}{4}) - \frac{\pi}{4} \right)^{\frac{2}{3}},
\end{equation}
it is again found that the apparatus matches the quantum mechanical prediction well as shown in Figure \ref{figure4}, though not as nicely as was the case for apparatus 1. 



\section{Discussion and conclusion}


With the help of apparatus 1, experimental data from CHSH type photon correlation experiments can extraordinarily well be imitated. Indeed, by the appropriate random insertion of further single counts and more often having the detectors output the 'wrong' bit, it should be possible to create data sets that are practically indistinguishable from actual experimental data sets, given current photon detector efficiencies and dark count rates (A Turing test for Bell correlation data may be a nice idea).



The argument presented here is not based on loopholes but a different interpretation of what happens in the experiment. Loopholes are related to possible flaws in the experiments such as communication between the two detectors, low detector efficiency or the mistreatment of data - in other words, loopholes question the validity of the \textit{outcomes} of experiments. Here, the position is that experiments provide a fair representation of the physical processes and that outcomes of the experiments are as such correct - given the experimental design choices. Neither is it argued that the quantum mechanical calculations so well matched by the experiments are mistaken or wrong. Both experiment and theory are believed to be correct. It is argued that the outcomes could be the result of the experimental design choice of incorporating a coincidence monitor or taking equal time measurements.

It may be nice to briefly digress to some wild speculation. Perhaps much of the discussion around the Bell inequalities has indirectly been a comparison between apples and oranges. Is it really so that the standard quantum mechanical calculation provides an event for event probability or could it be so that in essence it is an equal time calculation. If so, then may be there is no discrepancy between what is logically possible in a realistic and a quantum system. To drive the speculation and your patience to its limit, let me end this paragraph by conjecturing the following

\vspace{5mm}
\noindent
\textit{Conjecture:} \newline
\vspace{1mm}
\noindent
\textit{Bell correlations are the result of coincidence measurements.}
\vspace{5mm}

By using the coincidence monitor in the apparatuses, the question arises whether or not the simulations are truly non-local. Locality has been a key requirement for Bell experiments and even with a significant amount of imagination, it seems far-fetched though perhaps logically not entirely impossible, to drop this constraint. 

In the sense of how actual experiments are carried out, the apparatuses (in conceptual design identical to that of many experiments) are definitively local in the same way as the experiments of e.g.~Aspect and Zeilinger \cite{Aspect-1982,GHZ-1990}. There is clearly no mechanism for the left and right hand sides to communicate and no loopholes are exploited. 

However, it may not be entirely unreasonable to consider coincidence monitoring to be a non-local activity. After all, in coincidence monitoring a clear post-selection of data takes place. One could therefore argue that most if not all experiments to date are non-local in this sense. Whether or not one would take that viewpoint, however, in the discussion about whether or not a realistic theory is fundamentally possible, the important point is that experimental setup and apparatus exactly match as with regards to the imposed conditions. Since the apparatus is able to generate Bell correlations, it can therefore be concluded that the experiments to date do not substantiate the view that a hidden variable theory is impossible as strongly as hitherto believed.

Indeed, to exclude the kind of apparatuses presented here, in experiments, it may be necessary to calculate the correlations on an event by event basis and avoid the use of coincidence monitors or equal time measurements.




\begin{acknowledgments}
  I would like to thank Andreas Keil for valuable feedback on the manuscript.
  This work is partially supported by NUS Grant WBS: R-144-000-138-112.
\end{acknowledgments}
\bibliography{bcetm}

\end{document}